\documentclass[iop]{emulateapj}

\slugcomment{To appear in Astrophysical Journal. }

\shorttitle{Probing the active MBH candidate in NGC~404 with VLBI}
\shortauthors{Paragi et al.}

\begin{document}

\title{Probing the Active Massive Black Hole Candidate in the Center of NGC~404 with VLBI}

\author{Z. Paragi\altaffilmark{1}, S. Frey\altaffilmark{2}, P. Kaaret\altaffilmark{3}, D. Cseh\altaffilmark{4}, R. Overzier\altaffilmark{5}, 
P. Kharb\altaffilmark{6}} 

\altaffiltext{1}{Joint Institute for VLBI in Europe, Postbus 2, 7990~AA Dwingeloo, the Netherlands}
\altaffiltext{2}{F\"OMI Satellite Geodetic Observatory, P.O. Box 585, H-1592 Budapest, Hungary}
\altaffiltext{3}{Department of Physics and Astronomy, University of Iowa, Van Allen Hall, Iowa City, IA 52242, USA}
\altaffiltext{4}{Department of Astrophysics/IMAPP, Radboud University Nijmegen, PO Box 9010, NL-6500 GL Nijmegen, the Netherlands}
\altaffiltext{5}{Observat\'orio Nacional, Rua Jos\'e Cristino, 77. CEP 20921-400, Rio de Janeiro, Brazil}
\altaffiltext{6}{Indian Institute of Astrophysics, II Block, Koramangala, Bangalore 560034, India}

\begin{abstract}
Recently \citet{Nyland12} argued that the radio emission observed in the center of 
the dwarf galaxy \object{NGC 404} originates in a low-luminosity active galactic nucleus (LLAGN)
powered by a massive black hole (MBH, $M\la10^6$~M$_{\odot}$). High-resolution radio detections of MBHs
are rare. Here we present sensitive, contemporaneous {\it Chandra} X-ray, and very long baseline 
interferometry (VLBI) radio observations with the European VLBI Network (EVN).
The source is detected in the X-rays, and shows no long-term variability. 
If the hard X-ray source is powered by accretion, the apparent low accretion efficiency 
would be consistent with a black hole in the hard state. Hard state black holes are known to show 
radio emission compact on the milliarcsecond scales. However, 
the central region of NGC~404 is resolved out on 10~milliarcsecond (0.15--1.5 pc) scales. 
Our VLBI non-detection of a compact, partially self-absorbed radio core in NGC~404 implies that either the
black hole mass is smaller than $3^{+5}_{-2}\times10^5$~M$_{\odot}$, or the source does not follow
the fundamental plane of black hole activity relation. An alternative explanation is that the central 
black hole is not in the hard state.
The radio emission observed on arcsecond (tens of pc) scales may originate in nuclear 
star formation or extended emission due to AGN activity, although the latter would not be typical
considering the structural properties of low-ionization nuclear emission-line
region galaxies (LINERs) with confirmed nuclear activity.
\end{abstract}


\section{Introduction}

All massive galaxies are believed to host supermassive black holes (SMBH, $10^6-10^9$~M$_{\odot}$) 
in their centers \citep{Kor95}, and there is ample observational evidence for a strong link between
SMBH and host galaxy properties \citep{F&M00,Gebhardt00}. This in turn points to a single mechanism in SMBH
and host galaxy formation. For example, major galaxy mergers may lead to the formation of close pairs of 
binary black holes (BH) in the early Universe, that would eventually coalesce because of dynamical friction 
and, eventually, gravitational radiation \citep{Vol03,Vol06}. In hierarchical cosmological structure
formation models, galaxy formation can be traced back to the stage of hundreds of smaller merging components,
but how efficient BH growth was preceding this stage is currently not known \citep{Vol10}. An outstanding 
question is how the seed BH in the mass range $\sim10^2-10^5$ M$_{\odot}$ were formed, and what the 
properties and demographics of those seeds are \citep{Vol10}. 

In the future, gravitational wave detectors may detect bursts of gravitational radiation from 
coalescing BH binaries. But at present, the only way to study SMBH formation observationally is looking 
at active galactic nuclei (AGN) in various regimes of the electromagnetic spectrum. It may seem 
straightforward to study objects at very high redshifts, when the Universe was much younger.
The most extreme high-redshift quasars indeed show signs of violent BH--host 
galaxy growth at redshifts $z=6-8$ \citep[e.g.][]{Fan06,Mortlock11}, but for the most 
luminous quasars we have evidence that these have already grown central SMBH with masses 
$\sim10^9$~M$_{\odot}$ \citep{Barth03,Willott05}. The more typical galaxies
-- that may harbor lower-mass central BHs and/or show no AGN activity -- at those redhsifts 
are beyond the reach of current X-ray telescopes \citep{Willott11,Cowie12}. 
Luminous quasars at lower redshifts do not give a clue on SMBH-formation 
either, because during the growth process, the initial conditions of black hole seeds 
are quickly erased by an efficient accretion process and the merger history \citep{Vol06}. 
It is currently impossible to search for massive black holes (MBH)\footnote{ \citet{Nyland12} 
used the term intermediate-mass black hole (IMBH). This term is widely used in connection with 
non-nuclear massive black holes, as possible central engines for ultra-luminous X-ray sources
\citep[e.g.][]{Colbert06,Webb14}. In this paper we find the term MBH more appropriate, that 
describes scaled-down versions ($M\sim10^2-10^5$~M$_{\odot}$) of supermassive 
black holes in the center of galaxies.} in the $<10^6$~M$_{\odot}$ regime 
by resolving stellar kinematics in galaxies beyond the Local Group, 
but they can be detected if they show AGN activity \citep{G&H04}.
Detection of such MBHs would be very important to answer 
key astrophysical questions, such as the accretion efficiency in seed BHs. Since dwarf galaxies 
were much less affected by merger processes, MBHs in dwarfs may retain (to some extent) the 
original seed mass distribution as well \citep{vanWassenhove10}. 

The first two detections of MBHs with masses $\sim10^5$~M$_{\odot}$ were in the dwarf Seyfert 1 galaxy POX~52 
\citep{Filippenko03,Barth04} and in NGC~4395 \citep{Peterson05,Thornton08}. Since then there
have been other candidates reported: the latest study by \citet{Reines13} included 151 dwarf galaxies
where they found 10 well-established broad-line AGN candidates (with AGN or composite spectra) with virial
BH masses\footnote{Assuming that the gas in the broad-line region is virialized, the luminosity and the 
width of the broad H$\alpha$ emission line can be used to infer the average gas velocity, and a scaling
relation can be applied to estimate the BH mass \citep[e.g.][]{G&H05,Reines13}.} 
in the range  $10^5-10^6$~M$_{\odot}$. It has to be noted that so far there are only
two cases confirmed with reverberation mapping, the above mentioned NGC~4395 \citep{Peterson05}, 
and SDSS J114008.71$+$030711.4 \citep{Rafter11}. The radio regime provides an additional
source of information to confirm low-luminosity AGN (LLAGN) activity in these systems. 
The challenge is to distinguish between nuclear star formation, supernova remnant (SNR) complexes, 
and LLAGN interpretations using constraints on the brightness temperature, luminosity and spectral 
index probed by very long baseline interferometry (VLBI) \citep[see e.g.][]{Alexandroff12}. 
\citet{Alexandroff12} targeted local analogs ($z<0.3$) of Lyman Break Galaxies showing starburst-like 
or composite spectra, with the goal to find LLAGN activity. They indeed detected two MBH candidates 
in the mass range $10^5-10^7$~M$_{\odot}$ with the European VLBI Network (EVN).
Their detection rate was however quite low ($<30\%$), showing that this is already rather difficult 
at redshifts $z>0.1$. In the very local Universe ($d<19$~Mpc), \citet{Nagar02,Nagar05} observed
the full sample of confirmed LLAGN such as low-ionization nuclear emission-line region (LINER) and Seyfert galaxies, 
and found that more than half of these showed flat-spectrum radio emission compact on 0.15 arcsecond scales. 
These were all detected and showed compact structure on milliarcsecond (mas) scales as well. 
Recently \citet{Panessa13} summarized the sub-parsec properties of a complete sample of 28 Seyfert galaxies within
$d<23$~Mpc, based on data available from the literature and more recent observations for 14 sources \citep{Giroletti09, Bontempi12}.
Out of the 23 LLAGN that had radio emission detected on arcsecond scales, 17 were also detected with VLBI at
least at one frequency, and most of these showed resolved structure in a single or several components,
distributed on 10--100~mas scales. In many cases a significant fraction of the radio emission was resolved out.
The closest member of this sample is the lowest-mass MBH ever detected with VLBI,  
NGC~4395 at a distance of $d=2.6$~Mpc and with a dynamical mass of $\sim10^5$~M$_{\odot}$ 
\citep[][and references therein]{Wrobel06}.

Another candidate MBH is in the nearby dwarf galaxy \object{NGC 404} at a distance of just 
3.1~Mpc (1$\arcsec$ angular size corresponds to 15~pc projected linear size), 
which has a complex nuclear environment \citep{Nyland12}. 
This dwarf galaxy has been enriched by cold gas 
due to merger activity that occured about 1~Gyr ago \citep{delRio04, Bouchard10}.
Most of the cold H\,{\sc i} gas forms a doughnut shape with inner and outer radii $\sim$1--4~kpc
(few hundred arcseconds), and this region forms stars at a rate of 2.5$\times10^{-3}$~M$_{\odot}$/yr
\citep{Thilker10}. The presence of a nuclear star cluster within 0.7 arcseconds
(inner 10~pc) with a mass of $1.1\pm0.2\times10^7$~M$_{\odot}$ indicates recent nuclear star 
formation \citep{Bouchard10, Ravindranath01}, which is likely accompanied by a central MBH 
evidenced by a central light excess within the cluster at radii $<3$~pc, with a dynamical mass
estimate of $\sim4.5\times10^5$~M$_{\odot}$ (Seth et al. 2010). 
\citet{Binder11} detected two distinct X-ray sources with {\it Chandra}, a nuclear point source dominated
by a power-law, and off-nuclear diffuse emission best fit with a thermal model. Based on the observed photon 
index and the low Eddington ratio, they concluded that the properties of the hard X-ray source in NGC~404 
are consistent with an LLAGN. \citet{Nyland12} reported radio 
detection with the Karl G. Jansky Very Large Array (VLA) in the L-band (B-array) and C-band (A-array), 
with a 1.4-GHz flux density of 2.8~mJy, and steep spectrum ($\alpha=-0.88; S\propto\nu^\alpha$).
The position of the radio source is coincident with the optical nucleus and the hard X-ray source. 
In this paper we present sensitive EVN and {\it Chandra} observations of NGC~404, that simultaneously 
probe the hard X-ray and the sub-parsec radio source properties.
 
\section{Observations and Results}

\subsection{EVN and WSRT Observations}\label{EVN}

The EVN observations were carried out on 4 June 2013 from 02:15:00 (UTC) to 10:14:50 at 1.6~GHz
(1594.49--1722.49~MHz), using a total data rate of 1024~Mbit/s/telescope, dual circular polarization, 
and 2-bit sampling (see Table~\ref{EVNtels} for telescope parameters). 
The Westerbork Synthesis Radio Telescope (WSRT) was part of the VLBI network as a phased array, 
but it also produced local interferometer data
that were used to measure the source total flux density. NGC~404 was phase-referenced to the nearby
calibrator J0112+3522 (cycle time 5 minutes), separated by only 0.7 degrees. 0234+285 served as an 
additional amplitude calibration check source for the VLBI data, while for amplitude calibration of 
the WSRT interferometer data we observed a 10-minute scan on 3C48.  

The EVN data were analysed in the 31DEC13 version of the Astronomical Image Processing System 
\citep[AIPS, e.g.][]{van96} using standard techniques\footnote{http:/$\!$/www.evlbi.org/user\_guide/guide/userguide.html}.
We used the measured system temperatures and gaincurves for the a-priori amplitude calibration.
In the absence of measured system temperature data, we used nominal system equivalent flux densities (SEFD)
for Badary (330~Jy), Jodrell Bank Lovell Telescope (70~Jy), and Zelenchukskaya (300 Jy). The nominal SEFD
of 370~Jy provided too low amplitudes for Svetloe. We applied a scaling factor for Svetloe baselines to provide 
consistent amplitudes with other baselines using the compact and bright sources 0234+285 and J0112+3522.
We solved for inter-channel phase and delay differences first using 2 minutes of data on J0112+3522, then 
fringe-fitted the whole timerange to solve for the residual phases, delays and rates on the calibrators.
The solutions found for J0112+3522 were transferred to the target source as well, using two-point interpolation
with rate correction, to avoid errors due to phase-wrapping. After the initial amplitude and phase calibration,
we solved for the instrumental bandpasses in each channels using J0112+3522. We edited the data for radio 
frequency interference, and applied the calibration tables to all sources. There were no major problems seen
in the data, therefore we do not expect that the correlation losses in our experiment were larger than in a 
typical EVN observation at these frequencies. The data quality was generally very good. The phase-reference source 
was bright and very compact, with a separation from thetarget well within a degree.
The imaging was carried out in Difmap \citep{Shepherd94}, version 2.4e.
We have calibrated the WSRT local interferometer data in AIPS, also following standard 
procedures\footnote{https:/$\!$/www.astron.nl/radio-observatory/astronomers/analysis-wsrt-data/analysis-wsrt-data}. 
The primary amplitude calibration was done using 3C84. We then solved for the residual amplitude and phase solutions 
on the phase-reference source J0112+3522 and ran {\tt GETJY} to find the total flux density of J0112+3522 
as well as to calibrate its gain solutions. These solutions were then interpolated to the target source NGC~404.
Just like in case of the EVN data, imaging was carried out in Difmap.

\begin{center}

\begin{table}

\caption{European VLBI Network telescopes participating in the observations}
\label{EVNtels}
\begin{tabular}{lrr}
\noalign{\smallskip}
\hline
\hline
\noalign{\smallskip}
Radio telescope	&  Diameter (m) & SEFD$^{\rm a}$ (Jy)  \\ 
\noalign{\smallskip}
\hline
\noalign{\smallskip}
Effelsberg (Germany) & 100     & 19          \\
Badary (Russia)      & 32      & 330         \\
Jodrell Bank (United Kingdom) & 76      & 65          \\
Medicina (Italy)     & 32      & 700         \\
Nanshan (P.R. China) & 25      & 300         \\
Noto (Italy)         & 32      & 784         \\
Onsala (Sweden)      & 25      & 320         \\
Sheshan (P.R. China) & 25      & 670         \\
Svetloe (Russia)     & 32      & 360         \\
Toru\'n (Poland)     & 32      & 300         \\
Westerbork (Netherlands) & $13\times 25^{\rm b}$ & 40   \\
Zelenchukskaya (Russia)  & 32      & 300         \\
\noalign{\smallskip}
\hline
\noalign{\smallskip}
\end{tabular}

\begin{list}{}{}
\item[$^{\rm a}$] System Equivalent Flux Density 
\item[$^{\rm b}$] The telescope was used in phased array mode for the VLBI observations, but also produced local interferometer data. 13 telescopes of the 14-element array were used in the observations. \\
\end{list}

\end{table}

\end{center}

\begin{figure}
 \includegraphics[angle=0,scale=.45, bb= 28 163 567 704, clip]{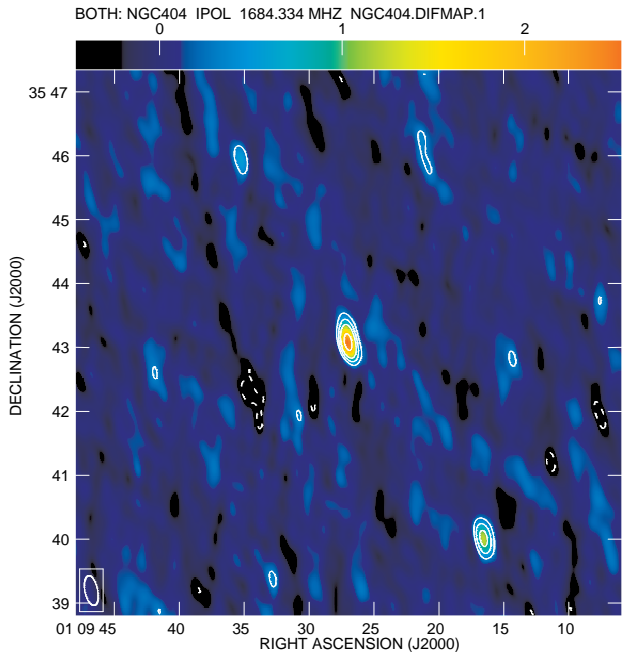}
 \caption{WSRT map of NGC~404 at 1.6~GHz (center). The peak brightness is 2.5~mJy/beam,
 the restoring beam is 28.4"$\times$11.9", with a major axis angle of 11.4$\degr$.
 The contour levels are at $\pm$300, 600, 1000 and 2000 $\mu$Jy/beam. 
 \label{fig_wsrt}}
 \end{figure}

The naturally weighted WSRT image shows NGC~404 as an unresolved source within the beam of 28.4$\times$11.9 
arcseconds with a position angle of $P\!A=11.4^{\circ}$ (see Fig.~\ref{fig_wsrt}). 
We fitted a point source model to the interferometer visibility data \citep[e.g.][]{Pearson95} that
resulted in a total flux density of 2.5$\pm$0.1~mJy (the error bar represents the rms noise in the map), 
in agreement with previous VLA results at this frequency \citep{Nyland12}. The high-resolution EVN data with a 
restoring beam of 14.0$\times$5.6~mas, $P\!A=3.0^{\circ}$ does not show a source, with an rms noise level of 
7~$\mu$Jy/beam.  
Within the 100 mas error circle of the source at position $\alpha_{\rm J2000} = 01^{\rm h}09^{\rm m}27.00^{\rm s}$
and $\delta_{\rm J2000} = +35\degr43'04.91"$ \citep{Nyland12}, the maximum peak brightness on the dirty map is 
22~$\mu$Jy/beam, and within 300 mas it is 29~$\mu$Jy/beam. Considering a 10--20\% correlation loss that is typical 
for EVN phase-referencing experiments at GHz frequencies \citep{Frey08}, we can rule out a compact radio source 
with a flux density of 35~$\mu$Jy (5$\sigma$ significance). 
Note that applying a Gaussian
$uv$-taper to decrease the resolution of the EVN data did not reveal any emission on larger ($\sim 30$~mas) scales either.
The radio emission is thus distributed on $>\!>$10 mas to a few hundred mas scales. This is somewhat surprising 
since the resolution of the 5--7~GHz VLA observations reported by \citet{Nyland12} was sub-arcsecond, and the 
source was at most moderately resolved. Nevertheless, our new data place a strong constraint on a 
compact component in NGC~404 on the 10-mas scales. We derive an upper limit to the brightness temperature,
defined as \citep{Kellermann88}: 
\begin{equation}
T_{\rm B}{\rm [K]} = 1.22\times(1+z) \frac{S}{\theta_{1}\theta_{2}\nu^{2}}
\end{equation}
where $S$ is the measured flux density in Jy, $z$ is the redshift (in this case zero), $\theta_{1}$ and $\theta_{2}$
are the major and minor axes of the beam in mas (or, for detected sources, the major and minor axes of the fitted
Gaussian components), and $\nu$ is the observing frequency in GHz. 
The 1.6~GHz brightness temperature upper limit is $\sim2\times10^5$~K
with the full resolution data, and $\sim2\times10^4$~K using the tapered data (resolution $\sim$30~mas, 5$\sigma$ 
upper limit 40~$\mu$Jy/beam). We note that Very Long Baseline Array (VLBA) observations of NGC~404 have been 
independently carried out at a similar resolution and somewhat less sensitivity compared to our data, with the
same result, i.e. the source was not detected (Kristina Nyland, priv. comm.).

\subsection{Chandra Observations}\label{Chandra}

We observed NGC~404 with the {\it Chandra X-ray Observatory} on 4 June 2013 from 02:24:14 (UTC) to 06:51:19, 
overlapping the time of the radio observations, and obtained a useful exposure of 13.6~ks, before 
deadtime correction.  
We also re-analyzed an archival observation obtained on 21 October 2010 04:03:55 
with an exposure of 98.3~ks, before deadtime correction\footnote{\citet{Binder11} 
quoted the exposure after deadtime correction.}, 
that was previously described by \citet{Binder11}. 
Both observations used the Advanced CCD Imaging Spectrometer (ACIS) with events recorded in VFAINT mode. 
We analyzed the standard level 2 data products using CIAO version 4.5 and CALDB 4.5.7 which retains the 
{\it Chandra} standard set of event grades equivalent to ASCA grades 0, 2, 3, 4, and 6 \citep{Fruscione06}. 
Neither observation was seriously affected by flaring.

NGC~404 was placed at the aimpoint of the ACIS-S3 chip in both observations.  
We extracted images of counts on the S3 chip in the 
0.3--8~keV and 2--10~keV bands and searched for sources using CIAO's {\tt wavdetect} tool. The nominal 
{\it Chandra} astrometric accuracy is $0.6\arcsec$ at 90\% confidence, but comparison of the positions of the 
four sources detected in both observations shows an offset of $1.1\arcsec$ between the two observations. 
This is almost a factor of 2 larger than the typical uncertainty and may indicate an issue with the 
aspect reconstruction in one or both observations. However, an inspection of the aspect camera data, kindly
performed by Jean Connelly of the {\it Chandra} Aspect Operations team, shows no obvious reason to prefer 
the aspect solution of one observation versus the other. Thus, there is no reason to prefer the absolute
astrometry of either the 2013 or 2010 {\it Chandra} observation.

To attempt to improve the astrometry, we searched for optical counterparts to the X-ray sources and 
found an X-ray source in the 2013 observation at the location of Mirach ($\beta$~And), an M0III giant star. 
X-ray emission has been observed from several M-type giant stars, including HR 5512 that shows no 
indications of binarity \citep{Hunsch04}, so we identify the X-ray source with Mirach. This position was off 
the CCDs in the 2010 observation due to the different roll angle (both observations were pointed at NGC~404). 
Mirach has a large proper motion, $\sim 200$~mas/yr, and we calculate its position at the time of the 2013 
observation to be RA = 01$^{\rm h}$09$^{\rm m}$44.08$^{\rm s}$, DEC = +35\degr37'12.5". This matches the X-ray 
source position within $0.3\arcsec$.

\begin{figure}
 \includegraphics[angle=0,scale=.44, bb= 35 187 577 605, clip]{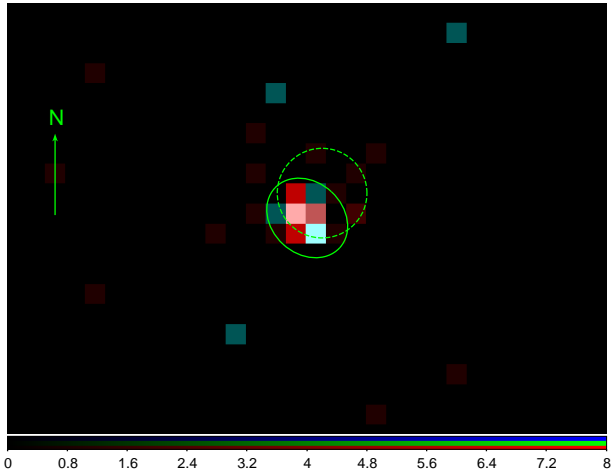}

\caption{{\it Chandra} detection of NGC~404. The figure shows X-ray counts from
the 2013 observation with no aspect correction in the soft (0.3--2~keV)
band in red and counts in the hard band (2--10~keV) in cyan. The green
dashed circle shows the position of the radio source reported by
Nyland et al. (2012) with a 1.1 arcsecond radius to indicate the uncertainty 
in the Chandra astrometry. The green ellipse shows the position of the hard
X-ray source. The arrow points north and has a length of 2 arcseconds.
\label{fig_Chandra}}
\end{figure}

We detect a hard X-ray source (in the 2--10~keV band) near NGC~404 in both the 2010 and 2013 
observations. After aligning the observations using the 4 other X-ray sources detected 
in both observation, 
the hard source position agrees within $0.42\arcsec$ between the two observations. This is consistent given 
the alignment accuracy and uncertainties on the individual source positions. In the 2013 observation 
(Fig.~\ref{fig_Chandra}), NGC~404 is detected at a significance of 
$18.8\sigma$ and with $47 \pm 7$ net counts\footnote{The reported significance of the detections
come directly from the {\it Chandra} tool {\tt wavdetect}. 
See http:/$\!$/cxc.harvard.edu/csc/columns/significance.html for an explanation.} in the 0.3--8~keV band and 
at $3.8\sigma$ significance and $7.7 \pm 2.8$ net counts in the 2--10~keV band. Without any aspect 
correction, the hard source position is RA = 01$^{\rm h}$09$^{\rm m}$27.03$^{\rm s}$, 
DEC = +35\degr43'04.3" (J2000). Aligning the astrometry of the 2013 observation to that of the 2010 observation 
moves the hard source by $0.9\arcsec$ to RA = 01$^{\rm h}$09$^{\rm m}$27.02$^{\rm s}$, DEC = +35\degr43'05.2" 
(J2000) which is consistent with the position reported by \citet{Binder11}. Due to the issues with the aspect 
solution, we conservatively assign an uncertainty of $1.1\arcsec$ (90\% confidence) to the hard source position.

The hard X-ray source had a count rate of $(5.7 \pm 2.1)\times 10^{-4}$~c/s in the 2--10~keV band in the 
2013 observation and a rate of $(5.8 \pm 0.8)\times 10^{-4}$~c/s in the 2010 observation. Thus, there is no 
evidence of a change in the average hard X-ray flux between 2010 and 2013. Assuming a power-law spectrum 
with a photon index of 1.9, we estimate a flux of 1.2 $\times 10^{{-14}}$ erg cm$^{-2}$ s$^{-1}$ in the 
2--10~keV band.

\subsection{Comparison with Previous Results}

In this paper we present new observational constraints with different instruments that 
have a wide range of angular resolutions, and therefore we probe different spatial scales 
in NGC~404. At the distance of NGC~404, 1" corresponds to 15~pc. The WSRT beam probes 
regions greater than 150 pc, {\it Chandra} has a spatial resolution of about 8 pc,
while the EVN beam roughly corresponds to 0.15 pc. It is also instructive to look at
the largest angular scales the EVN is sensitive to, to understand what spatial scales we 
cannot probe in the radio. The shortest baseline in units of observing wavelength was
1 million (1~M$\lambda$), and the second shortest was about 2~M$\lambda$. 
These would correspond to angular scales of 200~mas and 100~mas, respectively. 
Since imaging source structure requires more than a single baseline, 
the largest structure that can be detected with our VLBI array 
in the image plane is about 100~mas (corresponding to 1.5~pc). 
This means that we probe the radio emission between
0.15--1.5 pc linear scales with the EVN, but we do not have information on 
extended radio structures between 1.5 pc and 150 pc spatial scales.

The radio power of the arcsecond-scale radio source measured by the WSRT is 
$2.9\times10^{25}$~erg~s$^{-1}$~Hz$^{-1}$ ($2.9\times10^{18}$~W~Hz$^{-1}$) 
at 1.65~GHz. The total WSRT radio flux density of 
$2.5\pm0.1$~mJy at 1.65~GHz is fully consistent with the VLA 1.5-GHz flux density of 
$2.83\pm0.14$ given the steep spectral index \citep[$\alpha=-0.88$;][]{Nyland12}.
The observed 2-10~keV X-ray flux corresponds to an unabsorbed luminosity of 
$1.4 \times 10^{37}$ erg s$^{-1}$ at a distance of 3.1~Mpc. This is consistent with 
the luminosity estimated by \citet{Binder11} for the power-law component of the emission 
from the hard, nuclear point source using spectral fitting. 
It appears that the emission from within the central $\sim10$~pc did not vary significantly
since the previous measurements in the radio and in the X-ray bands.

The radio emission is resolved out with VLBI, indicating a source size 
exceeding 100~mas (1.5 pc). The 2-10~keV X-ray luminosities are usually compared 
to the $\nu L_{\nu}$ radio luminosities at $\nu=5$~GHz; by assuming a flat spectrum 
for a compact source (see next section), our strong upper limit of 35~$\mu$Jy 
corresponds to a 5-GHz radio luminosity upper limit of $\sim2\times10^{33}$~erg~s$^{-1}$ (or 
$\sim2\times10^{26}$~W) for the central 0.15--1.5~pc region. 

\section{Discussion}

\subsection{The Nature of the Nuclear Radio and X-ray Source in NGC~404}

\citet{Nyland12} considered various scenarios to explain the nuclear radio and X-ray emission: 
weakly accreting MBH, a stellar-mass X-ray binary system (XRB), a young SNR, 
and nuclear star formation. They have excluded the XRB and the young SNR scenarios based on
the observed radio and X-ray properties, and, in the latter case, also supported
by a statistical argument. They could not exclude the nuclear star formation scenario, but 
argued that the radio emission is most likely related to an active MBH. 
Our results of no long-term variability in the radio and in the X-rays, as well as the VLBI 
non-detection supports fully \citet{Nyland12} in that the radio emission is not related to an XRB 
or a young SNR, and may be consistent with nuclear star formation. Below we will consider what our
measurements imply for the active MBH scenario. 

In order to do that, we further calculate the ratio of the radio luminosity upper limit 
from the VLBI data and the X-ray luminosity. Using the definition for the radio-loudness parameter of 
$R_{\rm X}=\nu L_{\nu}$(5~GHz)$/L_{\rm X}$ \citep{Terashima03}, we find that log~$R_{\rm X}<-3.8$ for a 
compact source, in contrast to the earlier reported log~$R_{\rm X}=-2.5$ \citep{Nyland12}. The caveat
here is that {\it Chandra} and the EVN probe very different spatial scales in NGC~404. However, 
for hard-state MBHs in LINERs that show compact emission on 150~mas scales with the VLA,
most of the radio emission comes from a compact region detectable by VLBI as well; in some cases
these are resolved as pc-scale jets, but there are compact cores observed as well that are
interpreted as unresolved, partially synchrotron self-absorbed bases of jets\footnote{One cannot exclude that
in some cases the radio emission in unresolved radio cores is from the accretion flow \citep{Narayan00}.}
\citep[cf.][]{Nagar02,Nagar05}. Naturally, one cannot characterize the central black hole using
measurements that are dominated by extended emission from e.g. star formation. Our $R_{\rm X}$ value
is only correct under the assumption that all the X-rays come from a very compact region as well.
Currently the only way to assess this is looking at short-timescale variability in the X-ray emission.
Since \citet{Binder11} found evidence for such variability, we may conclude that at least a
reasonable fraction of the X-ray emission is from a very compact source.  
The Eddington ratio, defined as 
$\xi=log_{10}(L/L_{\rm Edd})$ ($L_{\rm Edd}=1.3\times10^{38} M_{\rm BH}/M_{\odot}$~erg\,s$^{-1}$)
can be inferred from the measured X-ray luminosity, assuming that it is 16\% of the bolometric 
luminosity \citep{Ho08}. In our case we obtain $\xi=1.5\times10^{-6}$ under the assumption that 
X-rays are fully powered by accretion and for an MBH mass $4.5\times10^5$~M$_{\odot}$ \citep{Seth10, Gultekin09}. 
This value is comparable to the Eddington ratio derived by \citet{Binder11} for the LLAGN case.

\subsection{The Case of an Accretion-Powered MBH}

A survey of the Palomar Spectroscopic Sample 
\citep{Ho97} at (mostly) 15~GHz with the VLA, and at 5~GHz with the VLBA 
showed that a very significant fraction of LLAGN have compact, flat-spectrum cores and 
(sub-)pc scale jets, indicating that at least 50\% of low-luminosity Seyfert galaxies and 
LINERs in the sample are accretion powered \citep{Nagar02, Nagar05}. NGC~404 was part of 
this sample but no  radio emission was detected with the VLA at the 10$\sigma$ levels of 
0.9, 1.3 and 10~mJy, at 8.4, 15, and 43~GHz, respectively \citep{Nagar00}. 
\citet{Nyland12} observed at lower frequencies (1.5, 5 and 7.5~GHz) and achieved significantly
lower noise levels. They detected slightly to moderately resolved radio structure from arcsecond 
to sub-arcsecond scales, with a spectral index of $\alpha=-0.88$ that is still broadly consistent 
with the properties of weakly-accreting LLAGN.  
The observed steep spectrum in AGNs is mostly indicative of optically thin synchrotron emission 
in extended structures, but there may be a faint, flat-spectrum compact source at lower 
flux density levels.
In addition, the radio source might be related to a single steep spectrum core, as 
observed in a number of LLAGN \citep[][to name a few]{Sadler95,Kharb10,OrientiPrieto10}. 
Because of the high angular resolution the VLBI technique provides, we can probe
if there is compact core emission as well as a (sub)-pc jet in NGC~404.

With our sensitive EVN data, we give a strong upper limit of 35~$\mu$Jy for compact emission. 
This is more than an order of magnitude deeper than the previous low-resolution 
measurement and upper limits indicated. Our non-detection does not automatically mean that 
an accretion-powered MBH can be completely ruled out in NGC~404, however, it is in contrast 
with the brighter population of LINERs that are known to have flat-spectrum cores and/or jets, and
the suggestion by \citet{Nagar05} that most LINERs might actually be powered by AGN activity. 
The estimated radio-loudness parameter log~$R_{\rm X} <-3.8$ is still 
consistent with a radio-loud object \citep[log~$R_{\rm X}>-4.5$,][]{Terashima03},
but note that we actually derive an upper limit. 
The Eddington ratio of $\sim1.5\times10^{-6}$ is somewhat lower, but close to the low end of what 
is implied in jet models of the radio emission in other LINERs  \citep[$10^{-1}-10^{-5}$,][]{Nagar05},
and significantly higher than the observed quiescent-state Eddington ratio of $\sim10^{-9}$ 
for Sagittarius~A*, that is well described by an 
advection-dominated accretion model \citep[see][for a review]{Feng14}.
Note that \citet{Nemmen13} recently studied a sample of 12 LINERs and found Eddington ratios
in the range $10^{-5}-5\times10^{-8}$. 

To compare our radio upper limit with weakly accreting black holes, we use the fundamental
plane of black hole activity \citep[FP,][]{Merloni03, Falcke04}. That is in principle a relation 
between the accretion efficiency and radio luminosity for a given mass. 
The latest and most accurate fit to the data including sub-Eddington black holes
was done by \citet{Plotkin12}. \citet{Miller-Jones12} performed a new regression of the
same data for determining black hole masses that takes the following form:

\begin{equation}
{\rm log} (M_{\rm BH}) = 1.638\, {\rm log} (L_{\rm R}) - 1.136\, {\rm log} (L_{\rm X}) - 6.863
\label{FP}
\end{equation}

\noindent where $L_{\rm X}$ is the 2--10~keV X-ray luminosity, $L_{\rm R}=\nu L_{\rm \nu}$ is the 5-GHz
radio luminosity (both in erg\,s$^{-1}$), and $M_{\rm BH}$ is the black hole mass in Solar masses. 
Using our measured X-ray luminosity, the 5$\sigma$ upper limit on the radio luminosity from a 
compact source, and assuming that NGC~404 is in a similar hard state like the sources that follow 
the FP-relation, we obtain a black hole mass upper limit of $3^{+5}_{-2}\times10^5$~M$_{\odot}$.
Here the error corresponds to the empirically determined uncertainty of 0.44 dex \citep{Miller-Jones12}. 

Here we stress that Eq.~\ref{FP} is  
particularly interesting when the radio measurement represents emission
from a compact source, i.e. it is obtained from a very high resolution instrument ($<\!<$arcsecond), 
or the 5~GHz flux density of a flat spectrum compact core is estimated from higher frequency measurements. 
The physics of the fundamental plane is well understood for massive and stellar-mass 
black holes in the hard state, in which case they are accreting significantly below their 
Eddington ratios. In this accretion state, the radio emission is dominated by partially synchrotron 
self-absorbed compact jets that have flat spectra and and show distinct structural properties on 
mas scales with VLBI \citep[see][and references therein]{Paragi13}. Sources that have steep spectra 
and/or show significant large-scale emission are less understood and are likely influenced more 
strongly by their environment. This makes black hole mass estimates from measured X-ray and radio 
luminosities less reliable.
There are two ways to identify compact (on mas scales), optically thick radio emission from partially 
synchrotron self-absorbed cores: by measuring a flat radio spectrum, or by directly constraining the 
size with VLBI measurements.

Is it still possible that the radio emission resolved out by VLBI is related to LLAGN activity? 
For example, in the case of the blue dwarf galaxy Henize~2-10, \citet{Reines11}
argued that the non-detection of the central radio source on mas scales does not 
necessarily rule out the presence of an accreting black hole, the radio emission may come from 
extended jets and/or lobes. In a following paper however \citet{ReinesDeller12} showed that in fact 
there is a compact radio source component as well, detected with a resolution of about 100 mas. 
This is similarly true for Seyfert galaxies with steep radio spectra, where most of the 
emission often comes from extended jets and lobes that are not compact on mas scales 
\citep{OrientiPrieto10, Panessa13}, but very compact emission is also present in several cases, 
especially in narrow-line Seyfert~1 galaxies \citep{Doi13}.
While both Seyferts and LINERs often show very compact radio emission, the former on average
have steeper radio spectra, and higher median bolometric Eddington ratios, although the contrast 
in the former case is large within the two subclasses: $L/L_{\rm Edd}=$$1.1\times10^{-3}$ for Seyfert~1s, 
$5.9\times10^{-6}$ for Seyfert~2s, and it is $1.0\times10^{-5}$ for LINER~1s, and $4.8\times10^{-6}$ 
for LINER~2s \citep[][and references therein]{Ho08}.
\citet{Nagar05} proposed that LINERs may be related to hard state black holes (low Eddington rate, capable
of producing jets), while Seyferts show soft state black hole activity (higher Eddington rate, no
observed compact jets, only very resolved large-scale jet emission seen); the observed Eddington ratios,
spectral indices and morphologies do not seem to fully justify such a simplistic division between the classes.  
In the case of NGC~404, our very high resolution measurements do not show evidence that this LINER
galaxy harbours a hard state black hole with compact jet emission. Under the assumption that the radio 
emission is related to hard state black hole activity and the source follows the FP-relation, the 
BH mass has to smaller than  $3^{+5}_{-2}\times10^5$~M$_{\odot}$. But we note that besides the well-known 
caveats of applying the FP-relation to estimate black hole masses, recent work has shown that even hard 
state black holes do not always obey the FP-relation \citep[see][and references therein]{HeinzMerloni13}.
Finally, it is possible that the active black hole in NGC~404 is not in the hard state and the extended
radio emission is due to jets/lobes. This can be probed by intermediate-resolution instruments.

\section{Conclusions}

We have shown that the radio emission in the central region of NGC~404 is
resolved out on mas scales, using very sensitive EVN observations.
The contemporaneous {\it Chandra} and EVN measurements allowed us to give 
a strong constraint on the mass of the central MBH under the assumption
that the central black holes is in the hard state and follows the fundamental plane
relation, 
without the additional uncertainty factor of possible variability between the epoch 
of the radio and the X-ray observations. We have also shown that there is
no long-term variability in the radio and X-ray bands.
 
While the detection of a compact radio source would have supported the claimed
accretion-powered MBH as the source of the radio emission, our VLBI non-detection 
is still consistent with a weakly accreting black hole. The extended radio 
emission may be due to nuclear star formation, or extended jets or lobes due
to LLAGN activity. 
A jet-lobe structure can be confirmed by radio observations that probe angular scales
of $100-1000$~mas, like e-MERLIN. In the future, the VLBI technique will remain
an important tool to probe LLAGN activity, but as our current example shows, 
additional short spacings of a few tens to a few 100~km are crucial as well to be
able to disentangle the various sources of the radio emission. In the near future,
the joint EVN-e-MERLIN array will be the best instrument to do so. We note that
with the Square Kilometer Array (especially SKA1-mid) it will be possible to give 
better constraints on nuclear MBH masses because of the great increase in sensitivity, 
but only when the array configuration and high-frequency coverage ($\nu > 3$~GHz)
is enough to provide significantly sub-arcsecond resolution.

\acknowledgments

We are very grateful to the anonymous referee for comments that improved our manuscript.  
ZP and SF acknowledge support from the Hungarian Scientific Research Fund (OTKA K104539).
ZP thanks for useful comments from Richard Plotkin and Marcello Giroletti. 
The EVN (\url{http://www.evlbi.org}) is a joint facility of European, Chinese, South African, 
and other radio astronomy institutes funded by their national research councils. The WSRT is 
operated by ASTRON (Netherlands Institute for Radio Astronomy) with support from the 
Netherlands Foundation for Scientific Research. The research leading to these results has 
received funding from the European Commission Seventh Framework Programme (FP/2007-2013) 
under grant agreement No. 283393 (RadioNet3). The scientific results reported in this article 
are based in part on observations made by the {\it Chandra X-ray Observatory}. We are grateful 
to Harvey Tananbaum who approved our DDT request to observe with Chandra during our EVN 
observing run. AIPS is produced and maintained by the National Radio Astronomy Observatory, 
a facility of the National Science Foundation operated under cooperative agreement by 
Associated Universities, Inc. 

{\it Facilities:} \facility{EVN}, \facility{Chandra}.

\end{document}